\documentclass[pdflatex,sn-mathphys-num]{sn-jnl}

\usepackage{graphicx}%
\usepackage{multirow}%
\usepackage{amsmath,amssymb,amsfonts}%
\usepackage{mathrsfs}%
\usepackage[title]{appendix}%
\usepackage{xcolor}%
\usepackage{textcomp}%
\usepackage{manyfoot}%
\usepackage{booktabs}%
\usepackage{algorithm}%
\usepackage{algorithmicx}%
\usepackage{algpseudocode}%
\usepackage{listings}%

\usepackage[italic=true]{derivative}
\usepackage{mathtools}
\usepackage{braket}

\newcommand{\eps}{\varepsilon}

\usepackage{float}
\newcounter{maintextfigure}

\begin{document}

\title[Article Title]{Effect of Population Imbalance on Vortex Mass in Superfluid Fermi Gases}


\author*[1]{\fnm{Lucas} \sur{Levrouw}}\email{lucas.levrouw@uantwerpen.be}

\author[2, 3]{\fnm{Hiromitsu} \sur{Takeuchi}}\email{takeuchi@omu.ac.jp}

\author[1]{\fnm{Jacques} \sur{Tempere}}\email{jacques.tempere@uantwerpen.be}

\affil[1]{\orgdiv{Theory of Quantum systems and Complex systems (TQC)}, \orgname{University of Antwerp}, \orgaddress{\street{Universiteitsplein 1}, \city{Antwerpen}, \postcode{2610}, 
\country{Belgium}}}

\affil[2]{\orgdiv{Nambu Yoichiro Institute of Theoretical and Experimental Physics (NITEP)}, \orgname{Osaka Metropolitan University}, \orgaddress{\street{3-3-138 Sugimoto}, \city{Osaka}, \postcode{558-8585}, 
\country{Japan}}}

\affil[3]{\orgdiv{Department of Physics}, \orgname{Osaka Metropolitan University}, \orgaddress{\street{3-3-138 Sugimoto}, \city{Osaka}, \postcode{558-8585}, 
\country{Japan}}}


\abstract{One of the fundamental parameters associated with quantized vortices in superfluids is the vortex mass, which is the inertia of a vortex. As of yet, this mass has not been observed in a superfluid. However, ultracold Fermi gases provide a promising platform in which recently much experimental progress  was made, offering tunability of the interaction as well as control on the single-vortex level. Not only can the scattering length be freely tuned, allowing exploration of the BEC--BCS crossover, but also an imbalance between different pseudospin states can be introduced. We study the effect of introducing this imbalance on the vortex mass, using a method based on an effective field theory for superfluid Fermi gases. We find that it is crucial to consider the imbalance in conjunction with nonzero temperatures; at some temperatures, the vortex mass is significantly enhanced while at others, the vortex mass is diminished. This pronounced temperature dependence highlights the need for careful tuning of experimental conditions and identifies favorable parameter regimes in which the vortex mass is likely to be observed.
 }




\maketitle

\section{Introduction}\label{sec1}

In population-imbalanced Fermi gases, the interplay between superfluidity and the presence of an excess component introduces new complexities to the dynamics of quantized vortices. By tuning the scattering length via Feshbach resonances, these gases enable a controlled exploration of the BEC--BCS crossover, while population imbalance provides a unique platform to study multi-component physics.
Recent experimental advances now enable the precise creation and manipulation of vortices within these systems \citep{kwon2021, delpace2022, hernandez-rajkov2024, grani2025}, sparking new interest in the long-standing question of a vortex mass, which is the effective inertia of a vortex \citep{levrouw2025pra, richaud2025}.
 The vortex mass is a crucial parameter in point-vortex models, which are effective descriptions that treat vortices as point-like objects moving in the velocity field generated by other vortices. While it has been extensively discussed in the context of bosonic two-component systems \citep{richaud2020, richaud2021, bellettini2023}, its behavior in population-imbalanced Fermi gases remains largely unexplored. Given that defining and measuring this quantity has historically proven difficult, understanding how population imbalance influences vortex inertia is essential for refining effective descriptions of superfluid dynamics.

Historically, different theoretical proposals have given rise to very different estimates for the vortex mass. \citet{popov1973} introduced the so-called relativistic vortex mass $\mathcal E/c^2$
in terms of the vortex energy per unit length $\mathcal{E}$ and the speed of sound $c$, an expression later rederived from the compressibility of the fluid \citep{duan1992,duan1994}. Because $\mathcal{E}$ depends logarithmically on the system size, so does the relativistic mass, implying a collective character. Therefore, we will refer to this as a \emph{global} vortex mass.
In contrast, different models predict \emph{local} masses which do not depend on the system size. \citet{baym1983} showed that the displaced superfluid gives rise to an \emph{associated mass}, which is a hydrodynamic contribution equal to the mass of the expelled fluid. Additional local contributions arise when matter or excitations occupy the core; we refer to this as \emph{internal mass}. In Bose mixtures, the minority species can accumulate there \citep{richaud2020, richaud2021, bellettini2023}, while in single-component systems, bound quasiparticles contribute, such as Caroli–de Gennes–Matricon (CdGM) states in Fermi gases that yield the Kopnin mass \citep{kopnin1978, kopnin1998}, or analogous states in Bose gases \citep{simula2018}.
In our recent work \citep{levrouw2025pra}, we considered the associated and internal masses in the context of an effective field theory (EFT) for superfluid Fermi gases. In contrast to previous literature, we found that these can also have a logarithmic system-size dependence, which implies they can be considered global.

Even though these theoretical predictions are very different, which is correct is still unclear from an experimental perspective. Experimental studies of vortex motion in superfluid helium and atomic Bose–Einstein condensates (BECs) have often achieved excellent agreement with massless point-vortex models \citep{donnelly1991, navarro2013, samson2016}. Also in superconductors a finite vortex mass was proposed \citep{suhl1965} and later observed in several experiments \citep{fil2007, golubchik2012, tesar2021, nakamura2024}, although reported values vary substantially across different platforms and measurement techniques.
The difficulty of detecting inertial effects in helium can be attributed to its extremely small vortex core size relative to other characteristic length scales, rendering such effects negligible. By contrast, in ultracold gases—where the core size can be comparable to other system scales—the vortex mass may play a more prominent role. For instance, experiments with solitonic vortices in elongated harmonic traps \cite{yefsah2013, ku2014} indicate a nonzero vortex mass, though their geometry complicates direct comparison with theoretical predictions for planar motion.
In (bosonic) ferromagnetic spinor condensates, it was reported theoretically and experimentally that the quantum Kelvin-Helmholtz instability generates eccentric fractional skyrmions \citep{takeuchi2022,huh2025}
 which can behave like massive quantum vortices \citep{kanjo2024, caldara2024}.
An as-of-yet unexplored next step for the case of Fermi gases is to extend the current experiments in which vortex trajectories can be observed  \citep{kwon2021, delpace2022, hernandez-rajkov2024, grani2025} to the case of population-imbalanced superfluid Fermi gases. Population imbalance was first realized in the pioneering experiments by \citet{zwierlein2006} and \citet{partridge2006} in harmonic traps and more recently in a box trap \citep{mukherjee2017}; in the latter case, experiments so far have been limited to highly imbalanced (non-superfluid) Fermi gases.
Extending these setups to include vortices would provide a way to probe the multi-component physics in this system.
Since Cooper pairing can only occur for pairs of different pseudospin, imbalance will lead to an excess density that is expected to localize at the core and behave similarly to the minority component in the case of two-component BECs. This localized polarization was studied before for a harmonic potential in the context of Bogoliubov--de Gennes (BdG) theory \citep{hu2007,takahashi2006}. One would expect that this occupation of the core would lead to an increase in the vortex mass, which is the topic of this paper.

In this paper, we will extend the approach we developed in \citep{levrouw2025pra} to the case of imbalanced Fermi gases. In Sec.~\ref{sec:theory}, we revisit our theoretical framework. We introduce the expressions for the vortex mass we will use and present ways to calculate these using an effective field theory for superfluid Fermi gases. We comment on the range of validity and limitations of this method, and how to include imbalance in this model.
We first calculate the vortex profiles of the superfluid and normal densities, which are discussed in Sec.~\ref{sec:profiles}. The results for the vortex mass are given in Sec.~\ref{sec:vortex-mass-results}.
We discuss the dependence on scattering length, imbalance and temperature. We comment on the implications for future experiments. 

\section{Vortex mass in the effective field theory framework} \label{sec:theory}

In our recent work \citep{levrouw2025pra}, we proposed a different way of calculating the vortex mass in the framework of an effective field theory for superfluid Fermi gases.
The starting point is to presume that the system consists of a superfluid and a normal component with densities $\rho_s$ and $\rho_n$.
Correspondingly, we consider the vortex mass to consist of two contributions $M_{\text{tot}} = M_a + M_i$ where
\begin{align}
    M_{a} &= 2\pi \int_0^\infty dr \, r\left(\rho_{s,\infty} - \rho_s(r)\right) \label{eq:associated-mass}\\
    M_{i} &= 2\pi \int_0^\infty dr \, r\left(\rho_n(r) - \rho_{n,\infty}\right),\label{eq:internal-mass}
\end{align}
where $\rho_{s,\infty}$ and $\rho_{n,\infty}$ are the bulk values of the superfluid and normal densities, of which the second one vanishes at temperature zero.
The associated mass $M_a$ equals the mass of the superfluid expelled from the core (per unit length) and corresponds to the associated or induced mass in classical hydrodynamics. The internal mass $M_i$ represents the mass of the normal component in excess of the background value. 

An advantage of this approach is that it can be used together with any model that can predict the radial dependence of the superfluid and normal densities, for example BdG theory. We will make use of an effective field theory (EFT) with imaginary-time action
\begin{multline} \label{eq:eft-action}
    S_{EFT}[\Phi,\bar\Phi] =\\
    \int_0^{\hbar \beta} d\tau \int d \mathbf{x} \Big[\hbar\frac{D(|\Phi|^2)}{2}\left(\bar \Phi\frac{\partial\Phi}{\partial\tau}- \frac{\partial\bar \Phi}{\partial\tau}\Phi\right) + \hbar^2 Q \frac{\partial \bar \Phi}{\partial \tau}  \frac{\partial \Phi}{\partial \tau}  - \hbar^2 R\left(\frac{\partial{|\Phi|^2}}{\partial \tau}\right)^2 \\
+ \frac{\hbar^2C}{2m} \left(\nabla \bar \Phi\cdot \nabla \Phi\right)
- \frac{\hbar^2 E}{2m} (\nabla |\Phi|^2)^2
+ \Omega_{sp}(|\Phi|^2)
\Big].
\end{multline}
where $\Phi = |\Phi| e^{iS}$ is the complex Bardeen--Cooper--Schlieffer (BCS) order parameter. The coefficients appearing in the action can be computed as a function of the scattering length $a_s$, the temperature $k_BT = \beta^{-1}$, the bulk order parameter $\Delta$, the average chemical potential $\mu = \frac{1}{2} (\mu_\uparrow + \mu_\downarrow)$ and the imbalance chemical potential $\zeta = \frac{1}{2}(\mu_\uparrow-\mu_\downarrow)$ using the expressions given in Appendix B of Ref.~\citep{levrouw2025pra}. The coefficients $D(|\Phi|^2)$ and $\Omega_{sp}(|\Phi|^2)$ are in addition dependent on the local order parameter. We use the imbalance chemical potential as the parameter that characterizes the population imbalance in the system. This can be seen as an effective Zeeman field for the pseudospin.
We then compute, using Fermi units, the average chemical potential $\mu/E_F$ and the bulk order parameter $\Delta/E_F$ as a function of $k_F a_s$, $T/T_F$ and $\zeta/E_F$ using the mean-field equations for the spatially uniform case, which we review in Appendix \ref{app:imbalance}. There we also show the mean-field phase diagram, consisting of superfluid and normal phases, as well as a region with phase separation. At temperature zero, the superfluid phase is for the most part unpolarized, leading to a perfectly-paired BCS superfluid. However, at interaction strengths $(k_F a_s)^{-1} > 1$, there also exists a spin-polarized superfluid state \citep{sheehy2006, parish2007, radzihovsky2010}. However, to investigate the stability of this phase, beyond-mean field effects would have to be included. Additionally, the Fulde--Ferrell--Larkin--Ovchinnikov (FFLO) phase has been proposed \citep{fulde1964, larkin1965}. At the mean-field level, however, it occupies only a very small region of the phase diagram and has not yet been observed experimentally. In light of these subtleties, we restrict our zero-temperature analysis to the unpolarized superfluid phase. We denote the corresponding critical imbalance chemical potential by $\zeta_c$. 
The mean-field equation of state is chosen for simplicity, but may be replaced by a different equation of state such as the Gaussian pair fluctuation (GPF) equation of state \citep{nozieres1985, hu2006}, which was already applied to imbalanced Fermi gases \cite{parish2007, klimin2012}.

The EFT was derived using a gradient expansion, assuming that fermionic degrees of freedom, varying over the pair correlation length, change more rapidly than the bosonic ones, characterized by the healing length \citep{lombardi2016}. Its validity, therefore, requires the pair correlation length to be much smaller than the healing length.
In the BCS regime, both lengths are comparable at low temperatures and scale as $ 1/\Delta$, diverging as $1/\Delta \propto \exp((k_F|a_s|)^{-1})$ in the BCS limit $(k_F a_s)^{-1} \to -\infty$ \citep{marini1998}. In the BEC limit $(k_F a_s)^{-1} \to +\infty$, the healing length diverges as $(k_F a_s)^{-1/2}$ while the pair correlation length vanishes. Near the critical temperature $T_c$, the healing length diverges as $(1-T/T_c)^{-1/2}$, whereas the correlation length remains finite \citep{palestini2014}.
Thus, the EFT is most accurate near $T_c$ or in the BEC regime. In current superfluid Fermi gas experiments \citep{kwon2021, delpace2022, hernandez-rajkov2024, grani2025}, typical parameters $(k_F a_s)^{-1} \gtrsim -0.5$ and $T/T_c \gtrsim 0.4$ are still within the regime of validity of the EFT. Outside this regime, we can extract mostly qualitative results; however, they may already provide a useful benchmark to theory and experiments. For conceptual simplicity, we will first address the zero-temperature limit, even if the theory is not quantatively accurate in this regime over the entire crossover. Afterward, we extend the calculation to finite temperatures.

We can express total, superfluid and normal densities as a function of the order parameter. For the total density, we can use the mean-field density function in the local density approximation (LDA)
\begin{equation} \label{eq:total-density}
\rho_{\text{tot}}(|\Phi|^2)
= m \int \frac{d\mathbf k}{(2 \pi)^3}\left[1-\frac{\xi_k}{E_k(|\Phi|^2)} X(E_k(|\Phi|^2)) \right],
\end{equation}
where $\xi_k = k^2/2m-\mu$ and $E_k(|\Phi|^2) = \sqrt{\xi_k^2 + |\Phi|^2}$ are the single-particle dispersion and Bogoliubov dispersion, respectively, and $X$ is given in terms of the Fermi--Dirac distribution $f_F$ as $X(\epsilon) = 1 - f_F(\epsilon+\zeta) - f_F(\epsilon-\zeta)$. We should note that this local density approximation misses some features of the vortex structure; most notably the Caroli--de Gennes--Matricon (CdGM) states \citep{caroli1964} are not accounted for.
The superfluid density is defined to be the phase stiffness and is given by
\begin{equation} \label{eq:superfluid-density-eft}
 \rho_s(|\Phi|^2) = 4mC |\Phi|^2.
\end{equation}
The normal density can then be computed as the difference between the total and superfluid densities
\begin{equation}
    \rho_{n}(|\Phi|^2) =  \rho_{\text{tot}}(|\Phi|^2) -  \rho_{s}(|\Phi|^2). 
\end{equation}
In the bulk, we can write the normal density as
\begin{equation} \label{eq:normal-density-eft}
    \rho_{n,\infty} 
    = \frac{2m}{3} \int \frac{d\mathbf k}{(2\pi)^3} \,k^2 Y(E_k(\Delta)),
\end{equation}
where $Y(\epsilon) = \partial X / \partial \epsilon = \beta [1+\cosh(\beta\eps) \cosh(\beta \zeta) ] /[\cosh(\beta\eps)+\cosh(\beta \zeta)]^2$.
This is consistent with mean-field results in the literature for the imbalanced superfluid  \citep{botelho2006, tempere2009}.
Also of interest is the imbalance density, which is the difference in occupation between the total densities of up and down components. As explained in Appendix~\ref{app:imbalance}, it can in the uniform case be computed from the thermodynamic potential $\Omega$ as $\Delta \rho = m \, \partial \Omega/\partial \zeta$. We choose to use the mean-field thermodynamic potential and compute the non-uniform imbalance density in the LDA approximation
\begin{equation}
\label{eq:imbalance-density}
\Delta\rho(|\Phi|^2) = \rho_{\text{tot},\uparrow}(|\Phi|^2) -\rho_{\text{tot},\downarrow}(|\Phi|^2) 
= m \int \frac{d\mathbf k}{(2 \pi)^3} \frac{\sinh(\beta \zeta)}{\cosh(\beta E_k(|\Phi|^2))+\cosh(\beta \zeta)}.
\end{equation}
In the unpolarized superfluid phase, the imbalance will vanish in the bulk. However, when introducing vortices, a finite polarization may appear, which is localized in the vortex core. This imbalance density consists of unpaired fermions.
 We note that the decomposition of the total density into superfluid and normal components is distinct from its decomposition into balanced and imbalanced parts, and the two splittings are not a priori related. That said, the imbalanced part can be regarded as belonging to the non-condensed part, which is related, but not equal, to the normal component; condensed fraction and superfluid fraction coincide in the BEC limit, but this is not the case in general.

\section{Vortex profiles} \label{sec:profiles}

To evaluate $M_a$ and $M_i$, we need to solve for radial profiles of the superfluid and normal densities.
For this, we first calculate the order parameter for a singly-quantized, radially symmetric vortex, which takes the form
\begin{equation} \label{eq:vortex-ansatz}
    \Phi(r,\varphi) = \Delta f(r) e^{i\varphi}
\end{equation}
in polar coordinates centered on the vortex, where $\Delta$ is the value of the order parameter in the bulk. The order parameter profile $f(r)$ can be obtained by solving the stationary equation of motion corresponding to the action \eqref{eq:eft-action}. As was derived in Ref.~\citep{levrouw2025pra}, the solutions have the following asymptotic behavior for large $r$
\begin{equation} \label{eq:asymptotic-order-param}
    f(r) \sim 1 - \frac{1}{4} \frac{\xi^2}{r^2}, 
\end{equation}
where the healing length $\xi$ is given in terms of the EFT coefficients as
\begin{equation}
\quad  \xi = \sqrt{\frac{\hbar^2}{m}\frac{C}{\Delta^2 G}}, \quad G = \left. \frac{\partial^2\Omega_{sp}}{(\partial|\Phi|^2)^2}\right|_{|\Phi|=\Delta}.
\end{equation}
In the unpolarized superfluid phase at zero temperature, the healing length does not depend on the imbalance chemical potential $\zeta$.
In Appendix \ref{app:num}, profiles computed from the stationary EFT equation of motion are compared for different values of the $s$-wave scattering strength $a_s$. Using the expressions 
\eqref{eq:total-density}--\eqref{eq:normal-density-eft} we can compute total, superfluid, and normal densities, as well as the imbalance density.
From Eq.~\eqref{eq:asymptotic-order-param} we can find the asymptotic behavior of the superfluid and normal density
\begin{align}
    \rho_s &= \rho_{s,\infty} \left(1 -\frac{1}{2}\frac{\xi^2}{r^2}\right) + O\left(\frac{\xi^4}{r^4}\right) \label{eq:asymptotic-superfluid} \\
    \rho_n &= \rho_{n,\infty} + \delta\rho_{n,\infty}\frac{1}{2}\frac{\xi^2}{r^2} + O\left(\frac{\xi^4}{r^4}\right). \label{eq:asymptotic-normal}
\end{align}
where we introduced $\rho_{s,\infty} = \rho_{s}(\Delta^2)$ and
\begin{equation}
    \delta\rho_{n,\infty} = - \Delta^2  \left.\pdv{\rho_n}{ |\Phi|^2}\right|_{|\Phi|=\Delta}
    =
    \rho_{s,\infty} - \Delta^2   \left.\pdv{\rho}{ |\Phi|^2}\right|_{|\Phi|=\Delta}.
\end{equation}
which can be computed from the EFT coefficients, as explained in Appendix~B of Ref.~\citep{levrouw2025pra}. 

\begin{figure}[ht]
\centering
\includegraphics[width=\textwidth]{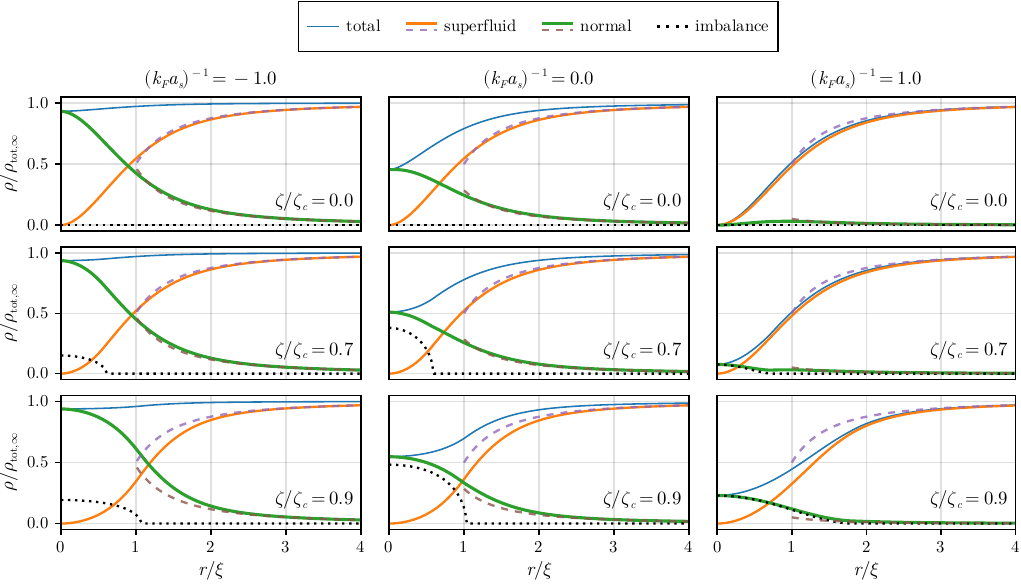}
\caption{This figure shows the total density $\rho_{\text{tot}}$ (solid blue line), superfluid density $\rho_s$ (solid orange line) and normal density $\rho_n$ (solid green line), as well as the density imbalance $\Delta \rho$ (dotted black line) as a function of the radial distance $r$ from the center of the vortex, calculated at temperature zero. All densities are scaled by the bulk total density $\rho_{\text{tot}\infty} = m k_F^3/3\pi^2$ and the radial distance is scaled by the healing length $\xi$.
These are shown for various values of the $s$-wave scattering length $a_s$ and imbalance chemical potential $\zeta$. Also plotted are the asymptotes of the superfluid density (dashed purple line) and the normal density (dashed brown line). }\label{fig:densities-T0}
\end{figure}

Profiles of the total, superfluid, normal and imbalance densities are plotted (for the zero-temperature case) in Fig.~\ref{fig:densities-T0}.
On the BEC side [$(k_F a_s)^{-1} = 1$] and at $\zeta = 0$, the normal density is only a small fraction of the total density, in particular vanishing in the origin. At finite imbalance, a normal density appears. This normal density mostly coincides with the imbalance density. The superfluid density is reduced in the core region (corresponding to a widening of the superfluid core), but it still has the same asymptotic behavior.
At unitarity [$(k_F a_s)^{-1} = 0$] and on the BCS side [$(k_F a_s)^{-1} = -1$], there is already a significant normal density present at zero imbalance. In this case, increasing the imbalance only slightly raises the normal density at $r=0$. Rather, the normal core becomes wider. Just as in the BEC case, the superfluid core is also expanded with respect to the case of zero imbalance.
In Appendix~\ref{app:num}, we quantitatively show the widening of the core by defining the vortex core radius and studying its dependence on the imbalance chemical potential.

\section{Vortex mass as a function of imbalance} \label{sec:vortex-mass-results}

Now, we can use Eqs. \eqref{eq:associated-mass} and \eqref{eq:internal-mass} to evaluate the internal and associated masses. As in the previous section, we begin by considering the zero-temperature case. Afterward, we discuss the dependence on temperature.

\subsection{Zero temperature}

As derived in Ref.~\citep{levrouw2025pra}, the associated and internal masses take the form
\begin{align}
    M_a &= \pi \xi^2 \rho_{s,\infty} \log\left(\frac{R}{\alpha_a \xi}\right) \\
    M_i &=
    \pi \xi^2  \delta\rho_{n,\infty} \log\left(\frac{R}{\alpha_i \xi}\right).
\end{align}
where $\alpha_a$ and $\alpha_i$ can be calculated by numerically evaluating the integrals \eqref{eq:associated-mass} and \eqref{eq:internal-mass}.
The logarithmic dependence on the system size follows directly from the asymptotic behavior of the superfluid and normal densities given in Eqs.~\eqref{eq:asymptotic-superfluid} and \eqref{eq:asymptotic-normal}. The prefactor sets the overal scale of the vortex mass. The correction factors $\alpha_a$ and $\alpha_i$ contain information about the vortex core.

\begin{figure}[ht]
\centering
\includegraphics[width=\textwidth]{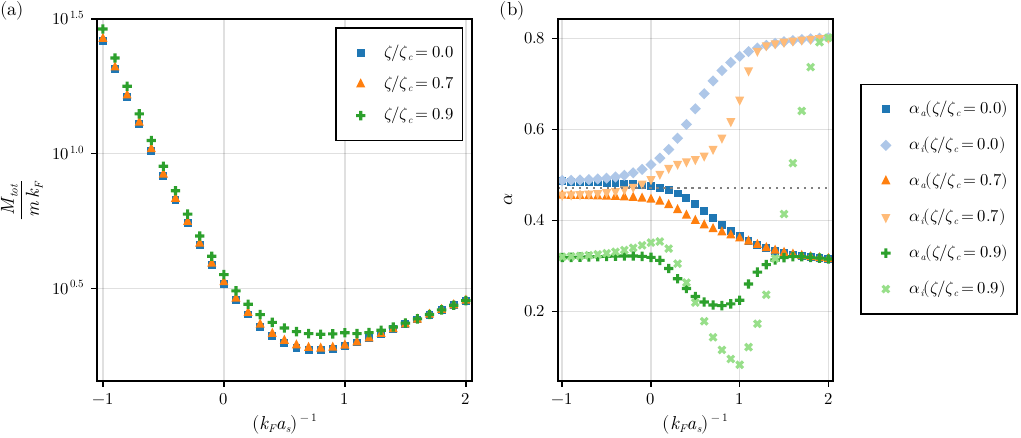}
\caption{(a) Total vortex mass for a system size $k_F R = 150$ as a function of the inverse scattering length and at zero temperature various values of the imbalance chemical potential $\zeta$.
(b) Correction factors $\alpha_a$ and $\alpha_i$ as a function of the inverse scattering length.}
\label{fig:vortex-mass}
\end{figure}

\begin{figure}[ht]
\centering
\includegraphics[width=\textwidth]{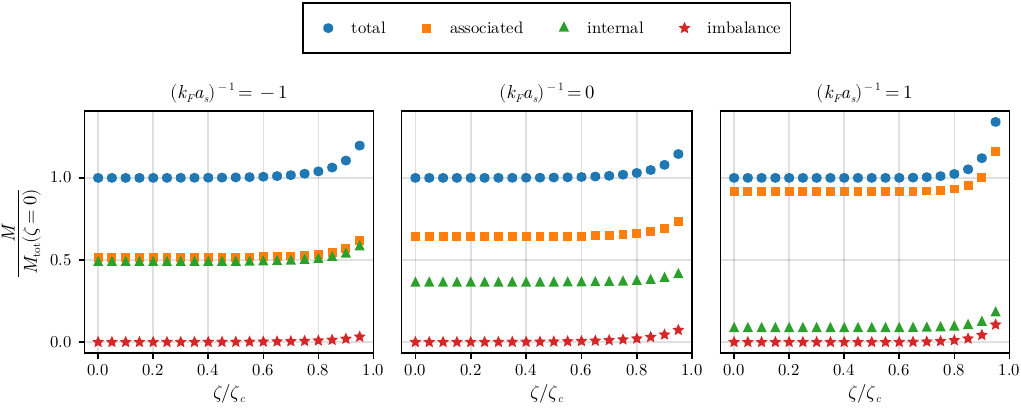}
\caption{Total, associated and internal vortex masses at zero temperature are given as a function of the imbalance chemical potential $\zeta$, for various values of the scattering length. Also the mass of the imbalanced component is shown. All masses are normalized by the total vortex mass at $\zeta = 0$.}
\label{fig:vortex-mass-zeta}
\end{figure}

The experiments \citep{kwon2021, delpace2022, hernandez-rajkov2024, grani2025} use a circular box potential with a radius of  $45 \,\mathrm{\mu m}$ and an inverse Fermi wave vector $k_F^{-1} \sim 0.3\,\mathrm{\mu m}$. Correspondingly, we evaluate the vortex mass at $k_F R = 150$. The results are shown as a function of the scattering length in Fig.~\ref{fig:vortex-mass}. Note that the cases in which there was a nonzero bulk imbalance density were not included.
Across the BEC--BCS crossover, we can see that the vortex mass increases when increasing the imbalance. However, as the imbalance only changes local quantities and does not affect the asymptotics of the vortex core, the increase is relatively small. 
The local correction to the vortex mass can be quantified by looking at the correction factors $\alpha_a$, $\alpha_i$. We can see that the imbalance has the effect of lowering these factors. Especially close to the critical imbalance, it is important to include this correction.
For all values of $\zeta$, we can see that $\alpha_i = \alpha_a$ in the BCS limit. This is the result of the observation that was made in Ref.~\citep{levrouw2025pra} that associated and internal masses are equal in this limit. In the deep BEC regime, $\alpha_a$ and $\alpha_i$ go back to their values at zero imbalance.
In Fig.~\ref{fig:vortex-mass-zeta}, the different contributions to the vortex mass are shown as a function of $\zeta$ for various values of the scattering length. In all cases, both the associated and internal masses increase. The first corresponds to the widening of the superfluid core; the second to extra quasiparticles located in the core. Both increases happen only close to the critical imbalance $\zeta_c$. This is also when there is a non-negligible amount of imbalanced component present. 

\subsection{Finite temperature} \label{ssec:finite-temperature}

\begin{figure}[ht]
\centering
\includegraphics[width=\textwidth]{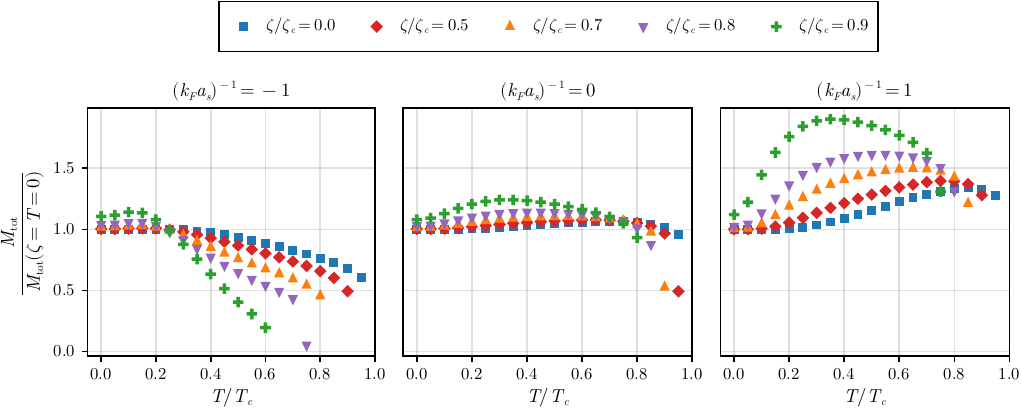}
\caption{Total vortex mass as a function of temperature at various values of the critical imbalance potential, given for inverse scattering lengths $-1$, 0 and 1. The critical imbalance chemical potential $\zeta_c$ is computed at temperature zero. All masses are normalized by the total vortex mass at $\zeta = T = 0$.}
\label{fig:vortex-mass-T}
\end{figure}

At finite temperatures, the normal density takes a finite value also in the bulk. Part of this normal density consists of the imbalance density. Profiles of the total, superfluid, normal and imbalance densities are given in Appendix~\ref{app:num}.
In Ref.~\citep{levrouw2025pra}, it was argued that, for the case of zero imbalance, higher temperatures lead to an enhancement of the vortex mass on the BEC side, but a decrease on the BCS side. This can be seen as a competition between two effects. On the one hand, the healing length increases with higher temperature, leading to a larger and more massive vortex core; on the other hand, as the background value of the superfluid density decreases and the one of the normal density increases, the profiles become shallower, leading to a lowering of the vortex mass.

The temperature dependence of the vortex mass is illustrated in Fig.~\ref{fig:vortex-mass-T}. 
In the BCS regime $[(k_F a_s)^{-1} = -1]$, the vortex mass generally decreases with temperature. 
However, near the critical imbalance ($\zeta/\zeta_c = 0.9$), a slight initial enhancement is visible before the mass drops sharply. 
At unitarity $[(k_F a_s)^{-1} = 0]$ and in the BEC regime $[(k_F a_s)^{-1} = 1]$, the mass exhibits a more prominent non-monotonic behavior, first increasing with temperature before decreasing. 
This behavior is strongly amplified by the population imbalance. 
Most notably, in the BEC regime with $\zeta/\zeta_c = 0.9$, the vortex mass almost doubles around $T/T_c \sim 0.35$. 
These results can be again understood as a competition between two effects. 
At finite temperatures, the healing length $\xi$ and background superfluid density $\rho_{s,\infty}$ become dependent on the imbalance chemical potential $\zeta$. As $\zeta$ is raised, $\xi$ increases and $\rho_{s,\infty}$ decreases. A figure showing the dependence of $\rho_{s\infty}$ and $\xi$ on imbalance and temperature is provided in Appendix~\ref{app:num}. At lower temperatures, the first effect dominates, leading to an increase in the vortex mass, while at high imbalance, the second effect is more prominent.
This results in a significant interplay between population imbalance and thermal fluctuations in determining the vortex mass.

\section{Conclusion and Discussion}\label{sec13}

In this paper, we extended the calculation of the vortex mass developed in Ref.~\citep{levrouw2025pra} to the case of a population-imbalanced superfluid. To the best of our knowledge, the impact of imbalance on the vortex mass has not been considered in other theoretical approaches.
We made use of an effective field description for superfluid Fermi gases along the BEC--BCS crossover, a framework that comes with an inherent limited range of validity. While we applied it across the full temperature and crossover regime, results at low temperature on the BCS side should be understood as qualitative rather than quantitative. Also, some effects are not accounted for, such as de CdGM states, which should give an extra local contribution which may lead to a quantitative correction on the BCS side. However, as argued in Ref.~\citep{levrouw2025pra}, this would not change the order of magnitude of the vortex mass.

We found that to analyze the effect of imbalance, it is essential to consider the interplay with the temperature. 
At low temperatures, the vortex mass increases with imbalance. At temperature zero, this increase is only local and does not change the order of magnitude of the vortex mass. However, at small but nonzero temperatures, the vortex mass can be significantly increased. On the other hand, at larger temperatures, and particularly when approaching criticality, the vortex mass decreases as the imbalance is increased.
These estimates can guide future experiments toward parameter regimes in which it would be easier to observe the vortex mass. Moreover, the imbalance offers an additional tunable parameter, providing more stringent tests for theoretical models.

\backmatter

\bmhead{Supplementary information}

No supplementary files.

\bmhead{Acknowledgments}

L.L. and J.T. acknowledge financial support by the Research
Foundation Flanders (FWO), Projects No. G0AIY25N, No. G0A9F25N, and No. G060820N. H.T. is supported by JSPS KAKENHI Grants No. JP18KK0391 and No. JP20H01842;
and JST, PRESTO (Japan) Grant No. JPMJPR23O5.

\section*{Declarations}









\bmhead{Funding}
See \emph{Acknowledgments}.

\bmhead{Conflict of interest/Competing interests}
The authors have no relevant financial or non-financial interests to disclose.
\bmhead{Data availability}
The code used to generate the figures in this manuscript is available at the following
GitHub repository: \url{https://github.com/TQC-Antwerp/Vortex-Mass-in-Superfluid-Fermi-Gases}.

\setcounter{maintextfigure}{\value{figure}}

\begin{appendices}

\setcounter{figure}{\value{maintextfigure}}

\makeatletter
\@addtoreset{figure}{section} 
\makeatother
\counterwithout{figure}{section}

\renewcommand{\thefigure}{\arabic{figure}}

\section{Mean-field theory of an imbalanced superfluid Fermi gas}\label{app:imbalance}
In the grand canonical ensemble, we consider the thermodynamical potential (per unit volume) $\Omega$ which is related to the Helmholtz free energy  (per unit volume) $\mathcal F$ by
\begin{equation}
    \Omega = \mathcal F - \mu_\uparrow n_\uparrow - \mu_\downarrow n_\downarrow = \mathcal F - \mu n -\zeta\, \Delta n,
\end{equation}
in terms of number densities $n_\sigma$ for the $\sigma$-component and writing $n = n_\uparrow+n_\downarrow$ and $\Delta n = n_\uparrow-n_\downarrow$.
In the mean-field approximation, the thermodynamic potential per unit volume $\Omega_{sp}$ can be calculated from the saddle-point value of the grand-canonical partition functional and is given by 
\citep{tempere2012}
\begin{multline}
    \Omega_{sp}(\Delta^2; \mu, \zeta, \beta)= -\frac{m}{4 \pi  \hbar^2 a_s}\Delta^2 \\
    -\int \frac{d \mathbf{k}}{(2 \pi)^3}\Big[ \frac{1}{\beta} \ln (2 \cosh \left(\beta E_k\right)+2\cosh (\beta \zeta))
    -\xi_k-\frac{m}{\hbar^2k^2} \Delta^2\Big],
\end{multline}
\begin{figure}[h]
    \centering
    \includegraphics[width=\textwidth]{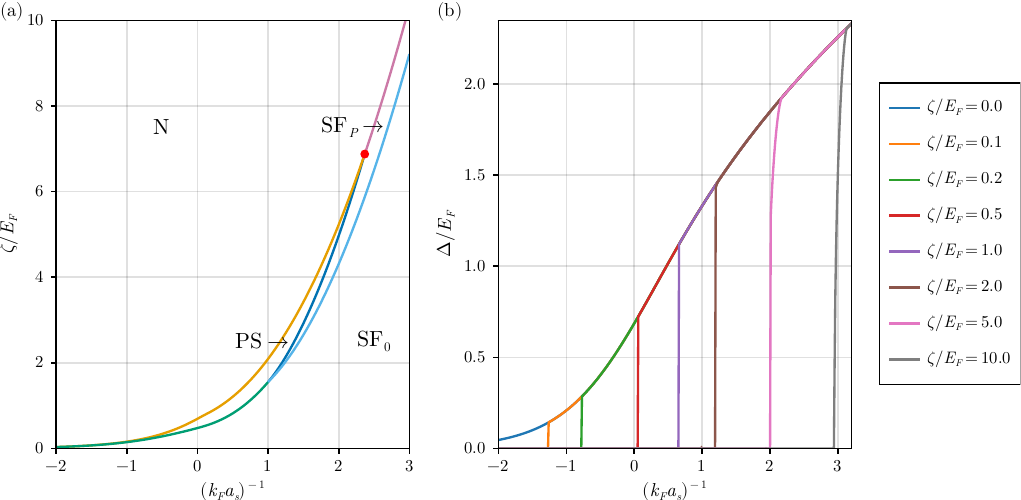}
    \caption{(a) Phase diagram for the Fermi superfluid at zero temperature. 
    We identify the normal phase (N), the unpolarized superfluid ($\text{SF}_0$), the polarized superfluid ($\text{SF}_P$) and the region of phase separation (PS). The tricritical point is shown in red. The critical imbalance chemical potentials $\zeta_c$ and $\zeta_c'$ as well as the minimum of the Bogoliubov dispersion are also plotted.  (b) The mean-field order parameter as a function of the inverse scattering length, for different values of the imbalance chemical potential $\zeta$. }
    \label{fig:zeta-saddlepoint}
\end{figure}
writing $\xi_k = k^2/2m-\mu$ and $E_k = \sqrt{\xi_k^2 + \Delta^2}$.
From this, the number densities can be computed by
\begin{align}
    n_{sp}(\Delta^2;\mu,\zeta,\beta) &= -\pdv{\Omega_{sp}}{\mu} = \int \frac{d\mathbf k}{(2 \pi)^3}\left[1-\frac{\xi_k}{E_k} X(E_k) \right], \\
    \Delta n_{sp}(\Delta^2;\mu,\zeta,\beta) &= -\pdv{\Omega_{sp}}{\zeta} =  \int \frac{d\mathbf k}{(2 \pi)^3} \frac{\sinh(\beta \zeta)}{\cosh(\beta E_k)+\cosh(\beta \zeta)} ,
\end{align}
where
\begin{equation}
    X(E_k) = \frac{\sinh(\beta E_k)}{\cosh(\beta E_k)+\cosh (\beta \zeta)}.
\end{equation}
The mass densities given in Eqs.~\eqref{eq:total-density} and \eqref{eq:imbalance-density} are defined as $\rho_{\text{tot}}(|\Phi|^2) = m\, n_{sp}(|\Phi|^2; \mu,\zeta,\beta)$ and $\Delta\rho_\text{tot}(|\Phi|^2) =m\, \Delta n_{sp}(|\Phi|^2; \mu,\zeta,\beta)$.
The mean-field values of the order parameter $\Delta$ and the chemical potential $\mu$ can be found by solving the saddle-point equations
\begin{align}
\frac{1}{a_s}&= - \frac{2}{\pi} \int_0^{\infty} d k\left[\frac{\hbar^2 k^2}{2m E_k}X(E_k)-1\right] \\
\frac{k_F^3}{3 \pi^2}&=\frac{1}{2 \pi^2} \int_0^{\infty} dk \, k^2\left[1- \frac{\xi_k}{E_k} X(E_k)\right].
\end{align}
 It is important to make sure that the obtained solutions are true minima of the thermodynamic potential. Carefully analyzing this leads to the phase diagram shown in Fig.~\ref{fig:zeta-saddlepoint}a. There are two superfluid phases. If $\zeta <  \min_{k} E_k = \sqrt{\min\{\mu,0\}^2 + \Delta^2}$, the superfluid is unpolarized (SF$_0$) i.e. the imbalance density is zero. In contrast, the imbalance density is nonzero in spin-polarized superfluid (SF$_P$). In the case the thermodynamic potential is lowest for $\Delta = 0$, we obtain the normal state. 
Also, there exists a phase-separated region; in this case, the system is unstable against phase-separation between the normal and superfluid state. The mean-field treatment can be extended to include an order parameter finite momentum, leading to the FFLO phase \citep{fulde1964, larkin1965}. However, this would only be stable in a very small region of the phase diagram.
In this manuscript, we limited our analysis at zero temperature to the unpolarized phase; therefore, we defined $\zeta_c$ to be the critical imbalance chemical potential to transition from the unpolarized superfluid to either the phase-separated state or the spin-polarized superfluid.
In Fig.~\ref{fig:zeta-saddlepoint}b, the zero-temperature solutions of the saddle-point equations are shown for various values of $\zeta$. The order parameter dropping to zero corresponds to entering either the normal or the phase-separated state.

\section{Numerical calculation of the vortex profile}\label{app:num}

\begin{figure}[t]
\centering
\includegraphics[width=\textwidth]{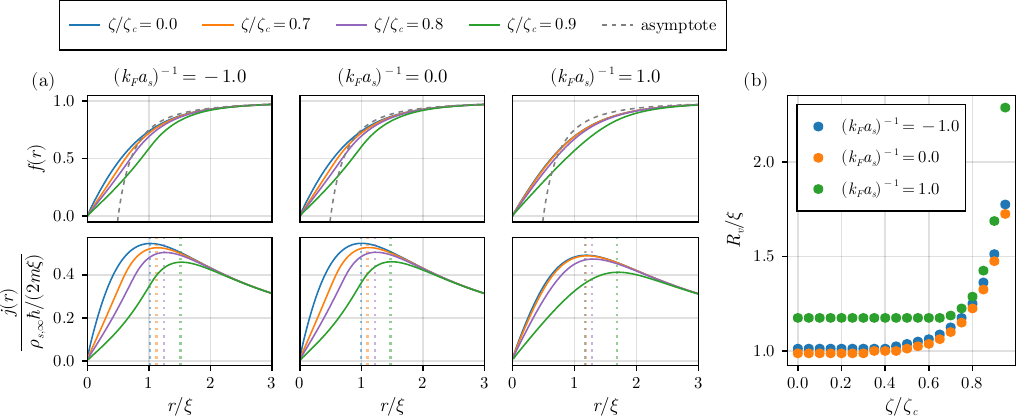}
\caption{(a) Order parameter profiles $f(r)$ and superfluid current densities $j(r)$ at various values of the $s$-wave scattering length $a_s$ and imbalance chemical potential $\zeta$, for temperature zero. At zero temperature, the asymptotic behavior does not depend on $\zeta$, except in the polarized superfluid phase. The dotted vertical lines show the maxima of the current, corresponding to the vortex core radius. (b) The vortex core radius $R_v$ as a function of  the imbalance chemical potential $\zeta$.}\label{fig:order-parameter-profile-T0}
\end{figure}

\begin{figure}[H]
\centering
\includegraphics[width=\textwidth]{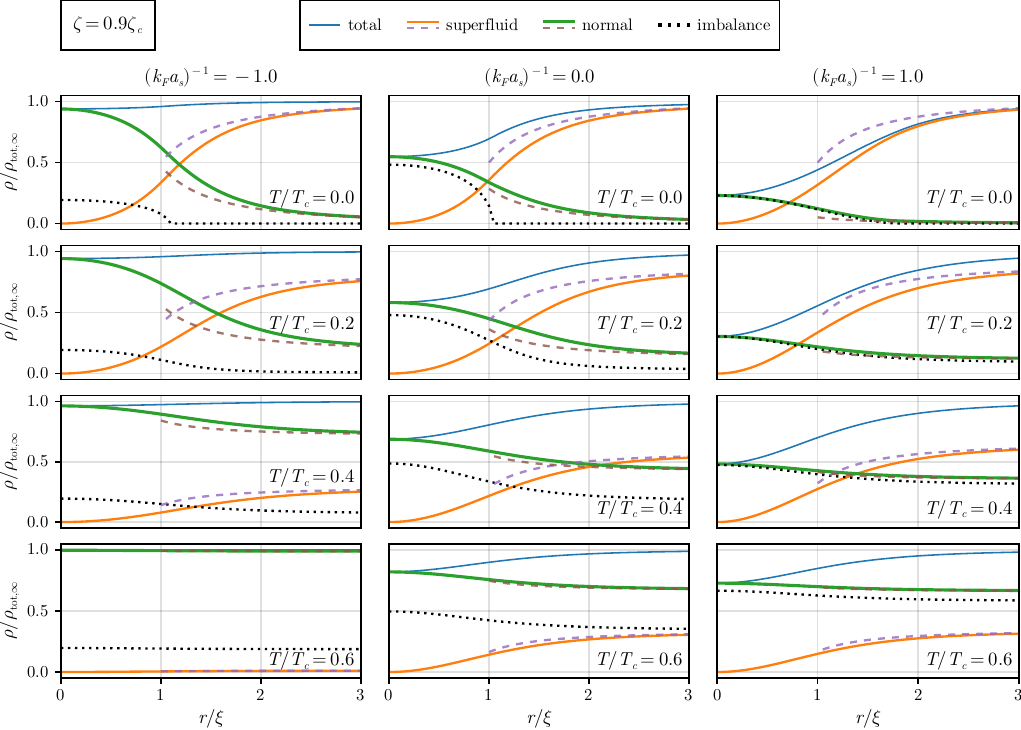}
\caption{This figure shows the total density $\rho_{\text{tot}}$ (solid blue line), superfluid density $\rho_s$ (solid orange line) and normal density $\rho_n$ (solid green line), as well as the density imbalance $\Delta \rho$ (dotted black line) as a function of the radial distance $r$ from the center of the vortex, calculated at an imbalance chemical potential $\zeta/\zeta_c = 0.9$. All densities are scaled by the bulk total density $\rho_{\text{tot}\infty} = m k_F^3/3\pi^2$ and the radial distance is scaled by the healing length $\xi$.
These are shown for various values of the $s$-wave scattering length $a_s$ and temperature $T/T_c$. Also plotted are the asymptotes of the superfluid density (dashed purple line) and the normal density (dashed brown line). }\label{fig:densities-finT}
\end{figure}

\begin{figure}[H]
\centering
\includegraphics[width=\textwidth]{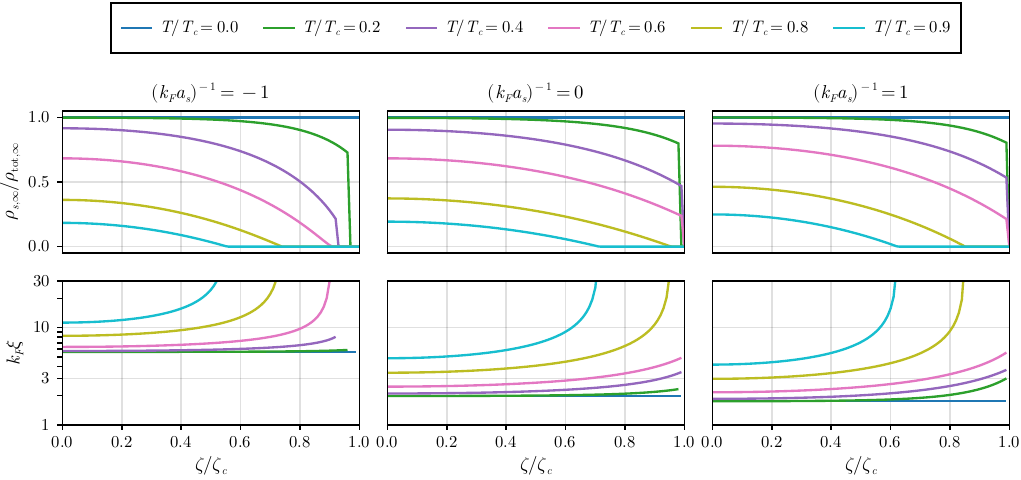}
\caption{Bulk superfluid density $\rho_{s\infty}$ and healing length $\xi$ are given in function of the imbalance chemical potential $\zeta$ and at various values of the temperature $T$.
}\label{fig:sf-densities-and-healing-lengths}
\end{figure}

The stationary equation of motion is
\begin{equation}
    -\frac{\hbar^2C}{2m} \nabla^2\Phi+\left( \mathcal A(|\Phi|^2) 
 + \frac{\hbar^2E}{m} \nabla^2 |\Phi|^2\right) \Phi= 0
\end{equation}
This equation was solved for the Ansatz \eqref{eq:vortex-ansatz} using the imaginary-time method described in Ref.~\citep{levrouw2025pra}. For this, a box of size $150\xi$ was used with a pixel size of $0.02\xi$. This leads to the profiles shown in Fig \ref{fig:order-parameter-profile-T0}a (at temperature zero). Also shown is (the magnitude of) the superfluid current density, which is given by
\begin{equation}
    j(r) = \rho_s(r) |\mathbf v_s(r)| = \rho_s(r) \frac{\hbar}{2m r}.
\end{equation}
The maximum of the current density can be identified as the vortex core radius $R_v$ \citep{simonucci2013}.
The vortex core radius as a function of the imbalance is given in Fig.~\ref{fig:order-parameter-profile-T0}b. From this, we can see that as the imbalance increases, the superfluid core widens, as was discussed in the main text.
From these, the density profiles given in Fig.~\ref{fig:densities-T0} were computed. Results at finite temperatures are presented in Fig.~\ref{fig:densities-finT}. 
As discussed in Sec.~\ref{ssec:finite-temperature}, the behavior of the vortex mass in function of temperature and imbalance can be understood in terms of the bulk superfluid density and healing length. These are shown in Fig.~\ref{fig:sf-densities-and-healing-lengths}.

\end{appendices}

\bibliography{references.bib}

\end{document}